\documentclass[%
 preprint, linenumbers,
 amsmath,amssymb,
 aps, physrev,
]{revtex4-2}

\usepackage{graphicx}
\usepackage{float}
\usepackage{dcolumn}
\usepackage{bm}
\usepackage{siunitx}
\DeclareMathOperator{\SO}{SO}
\DeclareMathOperator{\so}{\mathfrak{so}}

\DeclareMathOperator{\grad}{grad}

\begin{document}

\preprint{UHH}

\title{\textbf{Dynamics of a spherical ferromagnet on a non-conducting incline in an external magnetic field} 
}%

\author{Rasmus C. Raschke}
 \email{Contact author: rasmus.raschke@uni-hamburg.de}
\author{Tim Matthies}
\author{Roland Wiesendanger}
\author{Elena Y. Vedmedenko}
\affiliation{%
University of Hamburg\\
Physics Department\\
Jungiusstraße 11, 20355 Hamburg
}%

\date{\today}

\begin{abstract}
Magnetic particles of various shapes and sizes are used in biomedical applications, robotics, nanomagnetism, electrotechnology, and education. Yet despite years of study, a theoretical description of the rolling motion of a solid magnetic sphere still eludes us. One of the reasons is that coupling magnetic fields with mechanical degrees of freedom creates a highly non-linear motion with non-holonomic constraints, which makes modeling these systems analytically or by simulations non-trivial. The Lagrange-d'Alembertian approach to holonomic systems remains one of the most versatile methods to prove central theorems and solve problems in analytical mechanics and electrodynamics. While non-holonomic systems are not accessible by the original theory, recent developments generalize the principle to certain dissipative systems. \newline Here, we present a successful application of these developments to a simple mechanical system comprising a magnetic solid sphere on an incline within an external magnetic field. Using numerical simulations, we characterized how the system depends on its initial parameters. Furthermore, our findings are in line with recent experiments, demonstrating that even very weak magnetic fields can cause unexpected motion of ferromagnetic objects. The simplicity of this example also makes it a promising candidate for undergraduate physics education where the treatment of non-holonomic variational principles is still widely uncommon.
\end{abstract}

\maketitle

\section{INTRODUCTION}

The dynamics of rolling bodies on an incline form an essential part of the undergraduate physics syllabus. Despite being a well-known problem, even slight changes to the symmetry of the rolling body or the surface on which it is rolling can lead to interesting and unexpected dynamics, which are the subject of contemporary research. Prominent examples are provided by biased balls exhibiting various dynamic behaviors, including slow precession and nutation~\cite{Wheathead:AJP2020,Kilin:RCD2017}, surprising behavior of spherical magnets on an aluminum incline \cite{Cross:PE2025}, and dynamics of particle ensembles in viscous media \cite{Wolfschwenger:AMM2024}. The dynamics of objects with multiple rotational axes are particularly interesting. For instance, a biased ball that spins around its symmetry axis while also rotating around a vertical axis through its center of mass (COM) can exhibit complicated, curved trajectories of various shapes \cite{Jalali:PRE2015}. In most publications, both the rotational axes and their corresponding torques were considered to be of a mechanical nature. However, if a rolling body is magnetic, an additional, time-dependent magnetic torque and a corresponding axis appear in an external magnetic field. Consequently, even a perfectly shaped magnetic solid sphere could exhibit non-trivial rolling dynamics.

The first indication of this behavior was recently provided in  \cite{Vedmedenko:SciRep2022}. It has been demonstrated that if a magnetic solid sphere rolls due to some force and the magnetic and mechanical rolling torques are not collinear, the solid sphere's rolling trajectory will have a transversal component, causing it to start revolving and performing a type of planetary motion known as revolution. Although it has been pointed out in \cite{Vedmedenko:SciRep2022} that this revolutional motion depends on the relative orientation of the external magnetic field and the mechanical force, the shape of the trajectories remained unknown. 

The present investigation aims to provide a comprehensive description of the rolling dynamics of a ferromagnetic solid sphere on an incline. Our results demonstrate that the trajectory of such a sphere deviates significantly from linear rolling and exhibits revolution, with the center of the orbit lying outside the ball. However, the shape of the trajectory strongly depends on the initial mutual orientation of the magnetic moment $\hat{\bm{\mu}}$ and the external magnetic field $\bm{B}$. Interestingly, the trajectories change under exchange of $\hat{\bm{\mu}}$ and $\hat{\bm{B}}$. 
The dynamics of the rolling motion are complex and include spinning, precession, and nutation of the magnetic moment.

The manuscript is divided into the following sections. Section 2 is devoted to the analytical derivation of the equations of motion and a brief description of the numerical procedure, while chapter 3 comprehensively analyses the phase space of the rolling trajectories. Finally, we conclude by summarizing our findings.
\section{EQUATIONS OF MOTION}
\label{sec:theoretic}
\subsection{Analytical Derivation}
Consider a solid sphere of radius $R$ and mass $M$ rolling without slipping on a plane
\begin{equation}
    \label{eq:hypersurface}
    H_\alpha := \{ (x,y,z) \in \mathbb{R}^3 \mid h:=z-\alpha y =0 \}\subseteq \mathbb{R}^3
\end{equation}
of slope $\alpha$ and unit normal field $\bm{n}:=\frac{\grad h}{\|\grad h\|}$.
Assume that the solid sphere carries a magnetic dipole moment $\bm{\mu}_0$ interacting with a homogeneous, constant magnetic field $\bm{B}$. The configuration manifold is \[
Q := H_\alpha \times \SO(3) \subseteq \mathbb{R}^3 \times \SO(3)
,\] where $\SO(3)$ is the three-dimensional special orthogonal Lie group. The velocity phase space is thus given by \[
TQ = TH_\alpha \times T\SO(3) \cong TH_\alpha \times \so(3) \cong \mathbb{R}^2 \times \mathbb{R}^3 \subseteq \mathbb{R}^6,
\] where $T$ denotes the tangent bundle and $\so(3)$ is the Lie algebra of $\SO(3)$. We choose the trivial global chart on $\mathbb{R}^3$ and some local smooth chart $\phi: U \subseteq \SO(3) \to \mathbb{R}^3$ on $\SO(3)$. While constraining the motion to $H_\alpha$ reduces the system by one dimension, we choose not to eliminate the $z$ coordinate since we need the solution in the full frame for numerical analysis. A motion is then a function \[
(\bm{r}(t), P(t)): \mathbb{R} \supseteq I \to \mathbb{R}^3 \times \SO(3),
\] where $\bm{r}(t)$ is a path in $\mathbb{R}^3$ and $P(t)$ is a path in $\SO(3)$, i.e., a time-parametrized special orthogonal matrix. We define $\bm{\theta}:= \phi \circ P: I \to \mathbb{R}^3$. Our goal is to find equations of motion and their solutions \[
(\bm{r}(t),\bm{\theta}(t))=(x(t),y(t),z(t),\theta_x(t),\theta_y(t),\theta_z(t)): I \to \mathbb{R}^6
\] in local coordinates. Using the canonical Euclidean basis $\mathcal{B}:=(\bm{e}_x,\bm{e}_y, \bm{e}_z)$, we define a non-moving (inertial) coordinate frame by choosing an arbitrary origin. Furthermore, we consider a rotating coordinate frame defined by $\hat{\mathcal{B}}(t):=P(t)\mathcal{B}$. The angular velocity in the inertial frame $\bm{\Omega}(t)=(\Omega_x, \Omega_y, \Omega_z)^\top$ is defined in terms of the cross product matrix $$\Omega^\times(t) = \dot{P}(t)P^\top(t) = 
\begin{pmatrix}
    0 & -\Omega_z & \Omega_y \\
    \Omega_z & 0 & -\Omega_x \\
    -\Omega_y & \Omega_x & 0
\end{pmatrix}
.$$ 
\begin{figure}[htbp]
\centering
\includegraphics[width=\textwidth]{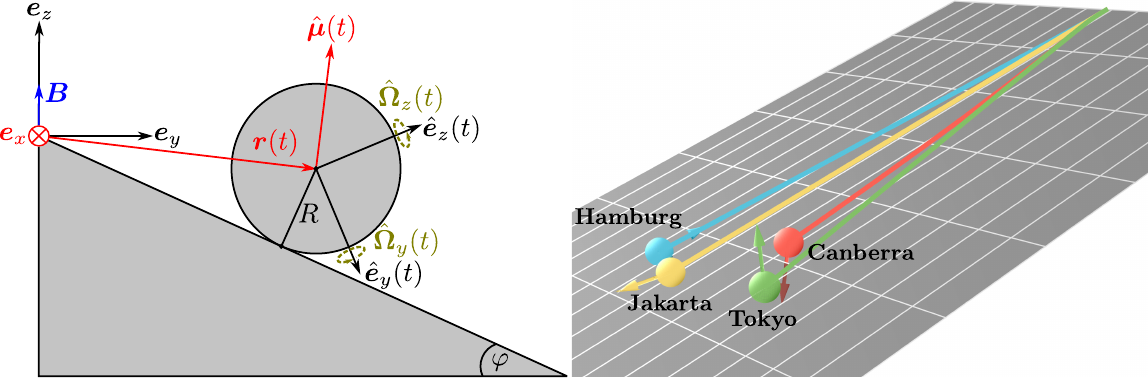}
\caption{Sketch of the system in the $yz$-plane and three-dimensional representation. The hat vectors are vectors of the basis $\hat{\mathcal{B}}$ while vectors without hat refer to $\mathcal{B}$. Red vectors are projected and may have non-zero $\bm{e}_x$-components. The plane $H_\alpha$ has an angle of inclination $\varphi$ defined by $\alpha = \tan \varphi$. The outer magnetic field is kept fixed parallel to $\bm{e}_z$. The three-dimensional representation shows four exemplary magnetic spheres moving along the incline at different points on Earth. The arrows indicate the direction of the magnetic moment $\hat{\bm{\mu}}$.}
\end{figure}
The kinetic energy form is given by 
\begin{equation}
    \label{eq:kinetic}
    T(\bm{r},\bm{\Omega})=\frac{M}{2}\|\dot{\bm{r}}\|^2 + \frac{I_0}{2}\|\bm{\Omega}\|^2
\end{equation}
and the potential by
\begin{equation}
    \label{eq:potential}
    U(z, \bm{\Omega})=-Mgz+\langle \hat{\bm{\mu}}(t), \bm{B}\rangle,
\end{equation}
where $\hat{\bm{\mu}}(t):=P(t)\bm{\mu}_0$, $I_0$ is the inertia scalar of an ideal ball, and $g$ is the gravitational acceleration. We obtain a Lagrangian $L(z,\bm{\Omega},\dot{\bm{r}}) := T - U$ and follow Ref.~\cite{Bloch2015} for non-holonomic systems. Assuming that there is an external force $\bm{F}_\text{e}=\bm{F}_\text{e}(\bm{r})$ such as friction, the corresponding Lagrange-d'Alembert principle reads \cite{Crepel2005, Lagrange1788}
\begin{equation}
    \label{eq:lagrangealembert}
    \delta \int_0^t L(z(s),\bm{\Omega}(s), \dot{\bm{r}}(s)) \, ds + \int_0^t \langle \bm{F}_\text{e}(\bm{r}(s)), \delta \bm{r}(s)  \rangle \, ds=0 
    .\end{equation} The torque can be written as \[
\bm{\tau}= \bm{\tau}_\text{em} + \bm{\tau}_\text{e},
\] where $\bm{\tau}_\text{em}:=\hat{\bm{\mu}}(t) \times \bm{B}$ is the magnetic torque, and $\bm{\tau}_\text{e} := -R \bm{n} \times \bm{F}_\text{e}$ is the torque induced by external forces.
The phase space of the translational motion admits an orthogonal decomposition $\mathbb{R}^3 \cong H_\| \oplus H_\perp$ with respect to the normal $\bm{n}$. Thus we can decompose arbitrary $\bm{v} \in \mathbb{R}^3$ as $\bm{v}=\bm{v}_\| + v_\perp \bm{n}$.
With the rolling constraint
\begin{equation}
    \label{eq:rollingconst}
    \dot{\bm{r}} = R \bm{\Omega} \times \bm{n},
\end{equation}
integration by parts, and non-degeneracy of the scalar product, one obtains tangential 
\begin{equation}
    \label{eq:tangangularvel}
    \frac{d \bm{\Omega}_\|}{dt} = \frac{1}{I_0+MR^2} \bm{\tau}_\| - \frac{RMg}{I_0+MR^2}(\bm{n} \times \bm{e}_z) 
\end{equation}
and normal 
\begin{equation}
    \label{eq:normalangularvel}
    \frac{d \Omega_\perp}{dt} = \frac{\tau_\perp}{I_0}
\end{equation}
equations for the angular acceleration from Eq.~\ref{eq:lagrangealembert}. Taking the derivative of Equation \ref{eq:rollingconst}, this fully determines the translational motion. Assuming Coulomb friction and aerodynamical drag by the drag equation, the external force becomes 
\begin{equation}
    \label{eq:friction}
    \bm{F}_\text{e} = - \nu_\text{r} Mg \langle \bm{n},\bm{e}_z\rangle \frac{\dot{\bm{r}}_\perp}{\|\dot{\bm{r}}_\perp\|} - \frac{\rho \, C A}{2} \|\dot{\bm{r}}\| \dot{\bm{r}}
    ,\end{equation} 
    where $\nu_\text{r}$ is the rolling resistance coefficient, $\dot{\bm{r}}_\perp$ is the tangential part of the velocity, $\rho$ is the fluid density of air, $C$ is the drag coefficient, and $A$ is the cross-sectional area of the sphere~\cite{Coulomb, Hayakawa2005, Landau2002}. For numerical reasons, we reparametrize $P(t)=P((q_0(t),\bm{q}(t)))$, where \[
(q_0(t), \bm{q}(t)): I \to \mathbb{S}^3
\] is a time-dependent unit quaternion. Subsuming, the full set of equations of motion are:
\begin{equation}
    \label{eq:eom}
    \begin{aligned}
        \frac{d^2 \bm{r}}{dt^2}&= \frac{5}{7M} \bm{\tau} \times \bm{n} - \frac{5g}{7} (\bm{e}_z)_\|, \\
        \frac{d\bm{\Omega}_\|}{dt} &= \frac{5}{7MR^2}\bm{\tau}_\| - \frac{5g}{7R} \bm{n} \times \bm{e}_z,\\
        \frac{d \Omega_\perp}{dt} &= \frac{5}{7MR^2} \tau_\perp, \\
        \frac{d(q_0,\bm{q})}{dt}&=\frac{1}{2}(0, \bm{\Omega}) \cdot (q_0, \bm{q}).
\end{aligned}
\end{equation}
\subsection{Numerical Solution}
The equations of motion were solved with our program $\mathtt{magsphere}$~\cite{code}. We implemented a Runge-Kutta-Munthe-Kaas-4 method, combining the quaternion-adapted Munthe-Kaas integrator with the Runge-Kutta 4 algorithm~\cite{Butcher1996, MuntheKaas1998}. To avoid numerical instability close to the singularity in Eq.~\ref{eq:friction}, we smoothed the directional function by
\begin{equation}
    \label{eq:smoothfriction}
        \frac{\dot{\bm{r}}}{\|\dot{\bm{r}}\|} \approx \tanh(k \| \dot{\bm{r}}\|) \frac{\dot{\bm{r}}}{\|\dot{\bm{r}}\|},
\end{equation}
where $k$ is a dimensionless constant describing the steepness of $\tanh$ around the origin. For the rolling resistance coefficient, we estimated $\nu_\text{r}=0.002$ based on values of $\nu_\text{r}$ for tires on steel and rubber surfaces~\cite{Guba2023}.
Assuming the International Standard Atmosphere and moderate velocities, values of $\rho = 1.2 \, \unit{kg.m^{-3}}$ and $C=0.47$ were used for simulations~\cite{iso2533, Baker1983}.
\section{RESULTS}
\subsection{Path Deflection in Earth's Magnetic Field}
\label{subsec:31}
We considered a ferromagnetic solid sphere of radius $R=0.5\,\unit{\centi\metre}$ and mass $M=5 \,\unit{\gram}$ carrying an initial magnetic moment $\bm{\mu}(0)=:\bm{\mu}_0$ with direction $\bm{\mu}_0 \| \bm{B}$ and magnitude $\|\bm{\mu}_0\|=:\mu_0=1.0\,\unit{\ampere\metre\squared}$ comparable to a real Neodymium magnet of this size~\cite{Amrani2015}. The solid sphere rolls on an incline with opening angle $\varphi=20^\circ$ in Earth's magnetic field $\bm{B}$. Data on $\bm{B}$ was obtained from the World Magnetic Model~\cite{ncei26}. Simulation time was set to $\Delta t=0.5\,\unit{\second}$ at a discretization of $\delta t=10^{-5} \,\unit{\second}$. The coordinate system is centered at the center of mass at $t=0 \, \unit{s}$ and the incline is rotated such that the $\bm{e}_x$-direction points eastwards, the $\bm{e}_y$-direction points northwards, and the $\bm{e}_z$-direction points up. The influence of the magnetic field was measured by the $x$-coordinate after $0.5 \, \unit{\second}$ because without an external field, the sphere's path would be contained in the $yz$-plane.
\begin{figure}[htbp]
    \centering
    \includegraphics[width=\textwidth]{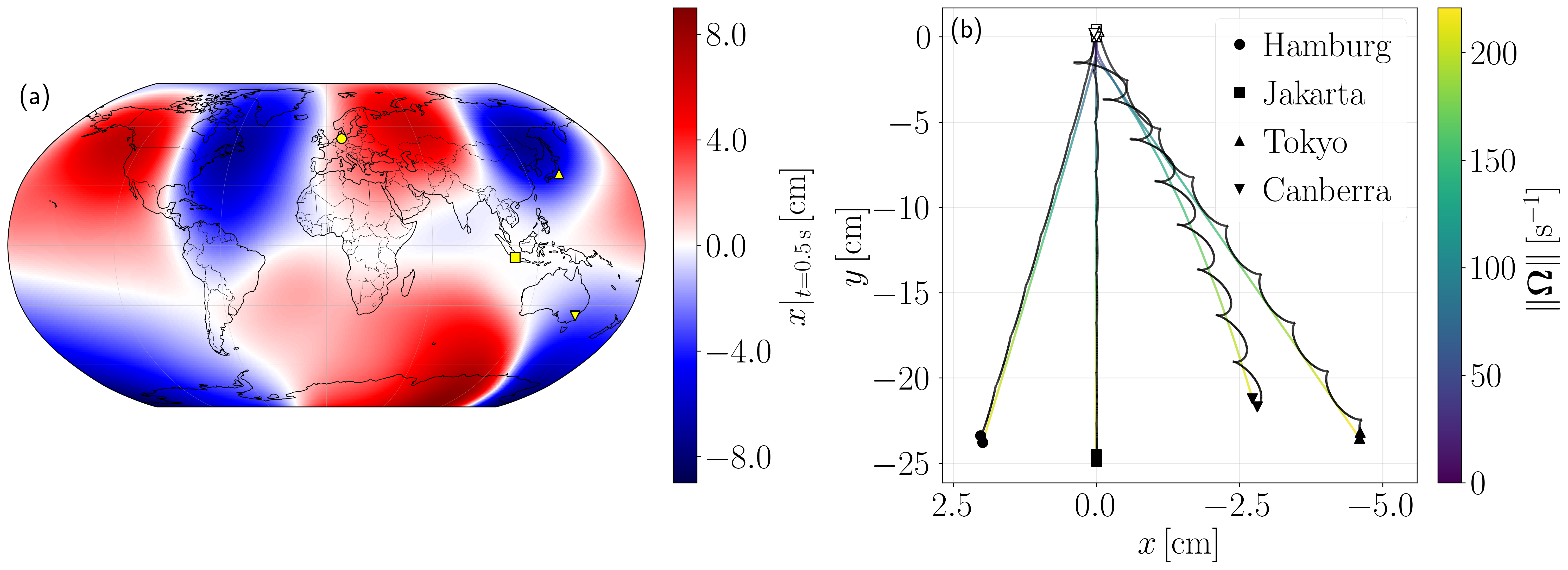}
    \caption{Robinson-projection of the deviation in $x$ depending on latitude and longitude (a) and selected trajectories in the $xy$-plane for different places on Earth (b). In (b), smooth lines represent the center-of-mass-trajectory colored by total angular velocity, while the black lines represent the projection of the direction of the magnetic moment vector normed to the sphere's radius. Initially, the magnetic moment $\bm \mu _0$ and the magnetic field $\bm B$ are aligned: $\bm \mu _0 \| \bm{B}$. The four trajectories are animated in Ref.~\cite{SM}.}
    \label{fig:k}
\end{figure} 
As seen in Fig. \ref{fig:k}, we observe a significant deflection of the solid sphere from an ideal straight path as it would be observed without any magnetic interaction. The strong dependence of the trajectories on the direction of the outer magnetic field leads us to predict that in a comparable experiment carried out in a laboratory, the results will vary noticeably with the location of the laboratory. 
\subsection{Influence of Magnetic Moment and External Field}
\begin{figure}[htbp]
    \centering
    \includegraphics[width=\textwidth]{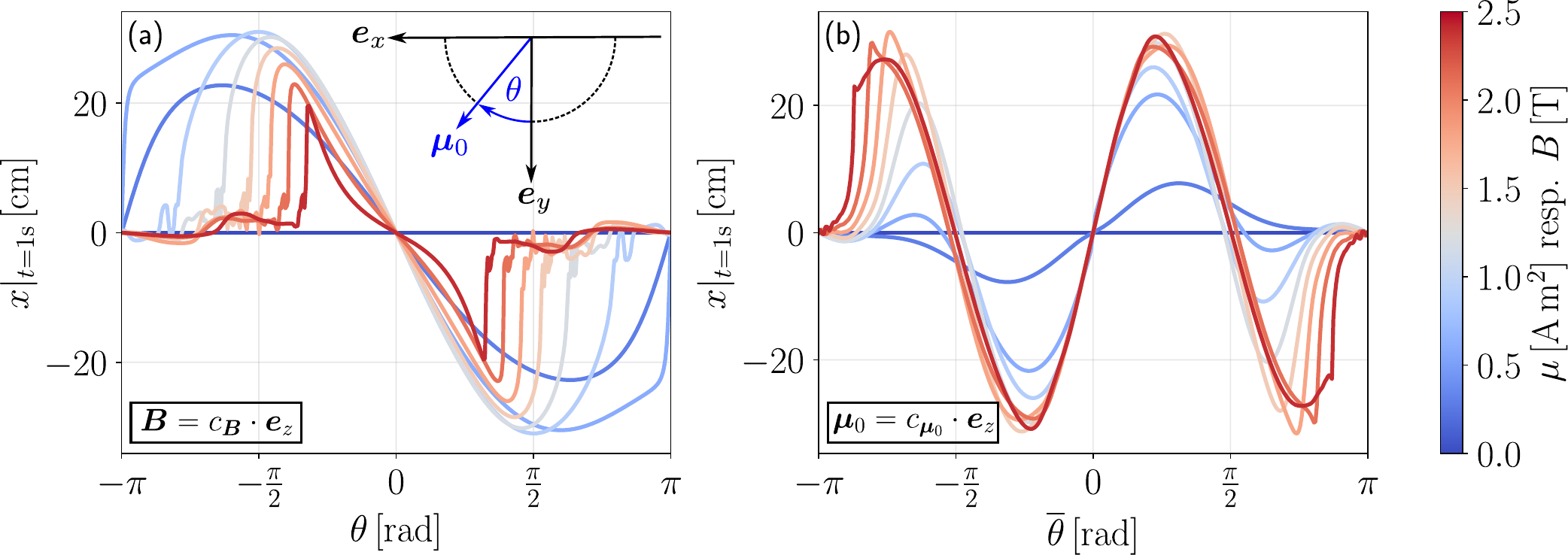}
    \caption{Final deviation $x|_{t=1\,\unit{s}}$ in relation to azimuthal angles $\theta, \overline{\theta}$ for different $\mu_0$ in (a) and different $B$ in (b). All graphs are point symmetric with respect to the origin. Critical points can be observed in both (a) and (b). Moreover, the critical points shift towards the origin and the maximal deviation increases with modulus in (a). The latter can also be seen in (b), yet the extrema do not follow a clear pattern for increasing $B$ like they do in (a).}
    \label{fig:f1}
\end{figure}

\begin{figure}[htbp]
    \centering
    \includegraphics[width=\textwidth]{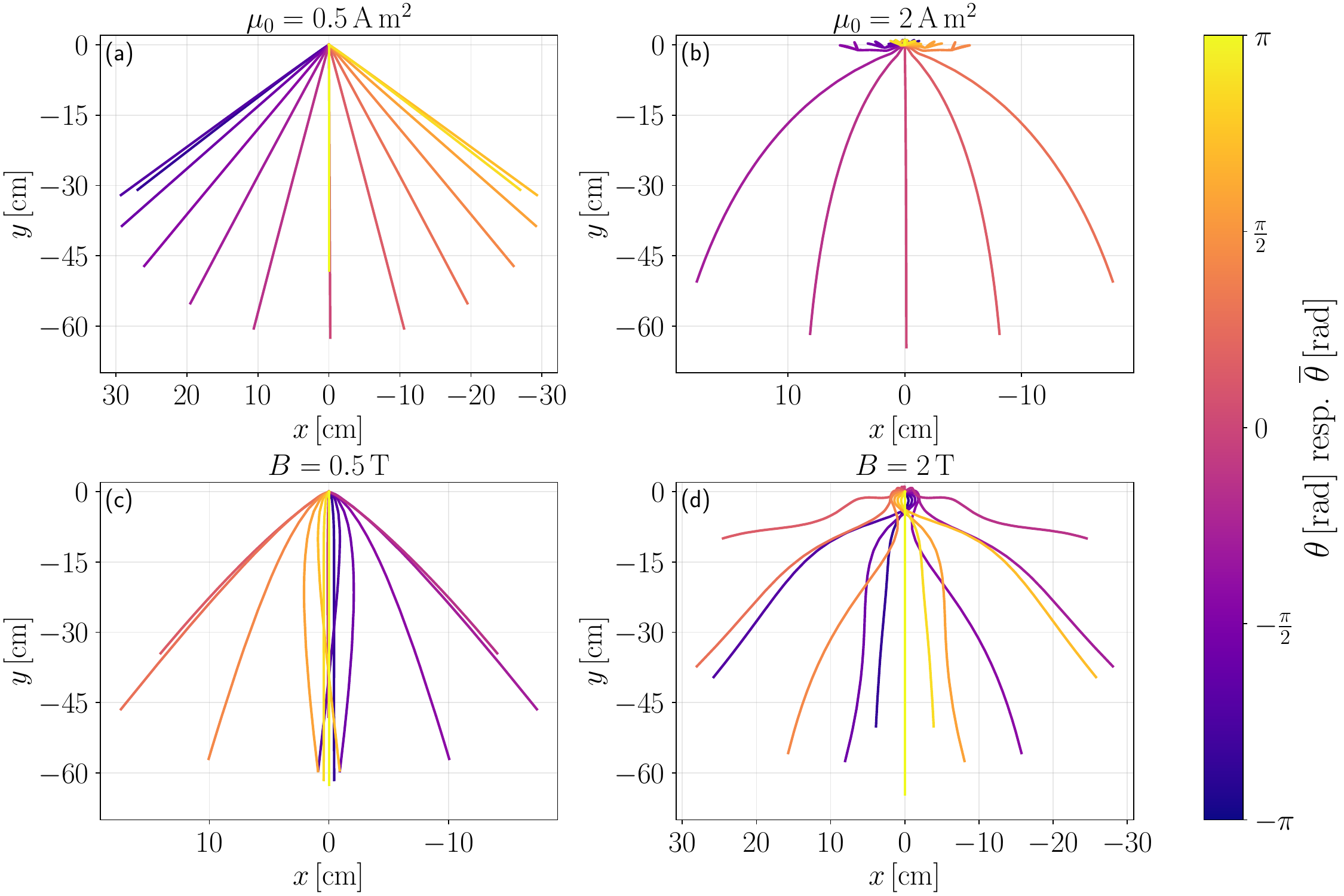}
    \caption{Trajectories on the $xy$-plane for selected moduli of $\bm{\mu}_0$ in (a,b) and $\bm{B}$ in (c,d). In (a) and (b), the curvature of trajectories increases with $\mu_0$ and $\theta$, yet the total deviation decreases for high values of $\theta$. In contrast, the paths in (b) start to loop around the origin for high values of $|\theta|$. In (c) and (d), the trajectories are strongly curved towards $x=0$, crossing over for high values of $|\overline{\theta}|$. Animations of exemplary trajectories are provided in Ref.~\cite{SM}.}
    \label{fig:f2}
\end{figure}

\begin{figure}[htbp]
    \centering
    \includegraphics[width=\textwidth]{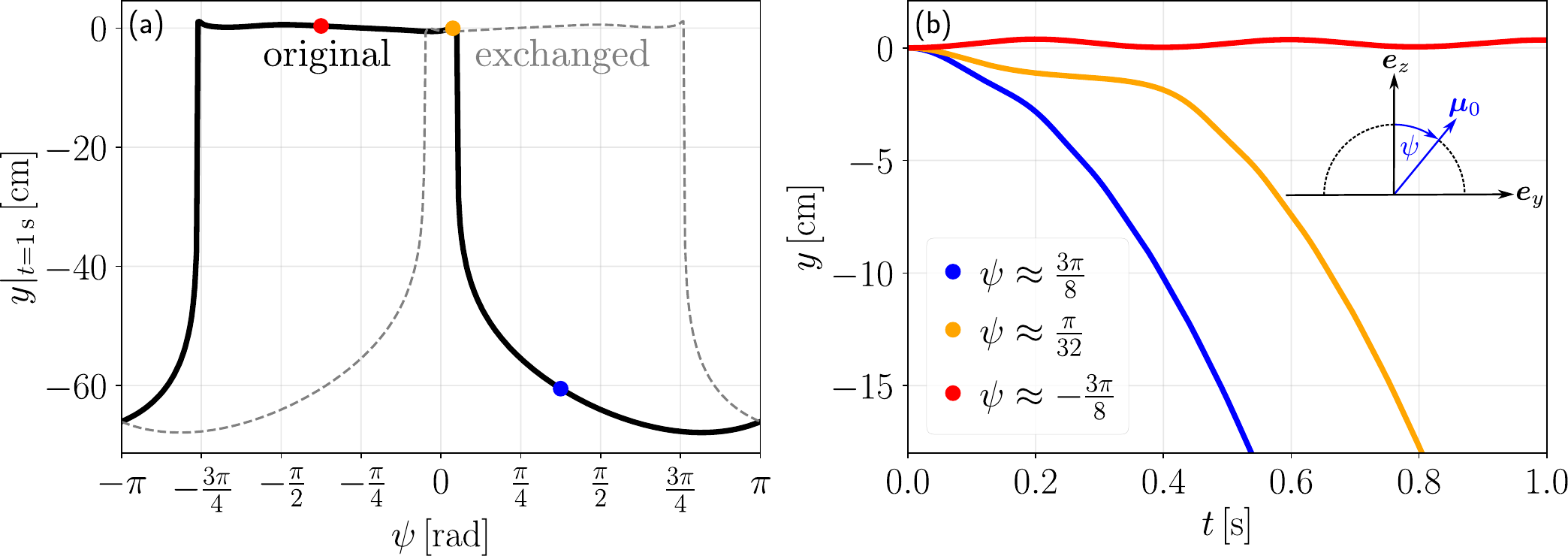}
    \caption{Final coordinate of the sphere $y|_{t=1\,\unit{s}}$ in relation to polar angles $\psi, \overline{\psi}$ in (a) and $y(t)$-curves for representative values of $\psi$ in (b). The grey dotted curve in (a) corresponds to exchanged roles of $\mu_0$ and $B$. In (a), three different regions are discernible: a high-deviation, a low-deviation, and an intermediate region. For each, a representative $y$-trajectory is shown in (b).  Exchanging $\bm{\mu}$ and $\bm{B}$ only causes a reflection of the curve in (a) around $\psi=0$. Videos of the three representative trajectories are contained in Ref.~\cite{SM}.}
    \label{fig:g}
\end{figure}

Aiming to understand the observed dependence of the system on the external field and the magnetic moment, we parametrized
\begin{equation}
    \bm{\mu}_0= \mu_0(\sin \psi \sin \theta, \sin \psi \cos \theta, \cos \psi)^\top
\end{equation}
and 
\begin{equation}
    \bm{B}=B(\sin \overline{\psi} \sin \overline{\theta}, \sin \overline{\psi} \cos \overline{\theta}, \cos \overline{\psi})^\top
\end{equation}
with azimuthal angles $\theta, \overline{\theta}$ and polar angles $\psi, \overline{\psi}$. First, we considered $\psi=\overline{\psi}=\frac{\pi}{2}$ and $\theta, \overline{\theta} \in \left[ -\pi, \pi\right]$, varying $\mu_0 \in [0.0 \,\unit{\ampere\metre\squared}, 2.5 \,\unit{\ampere\metre\squared}]$ while keeping $\bm{B}=c_{\bm{B}} \bm{e}_z$ fixed for $c_{\bm{B}}:=5 \cdot 10^{-5} \, \unit{\tesla}$. To examine the behavior under exchange of $\bm{B}$ and $\bm{\mu}_0$, we also varied $B \in [0.0\,\unit{\tesla}, 2.5 \,\unit{\tesla}]$ while keeping $\bm{\mu}_0=c_{\bm{\mu}_0}\bm{e}_z$ for $c_{\bm{\mu}_0}:=5 \cdot 10^{-5} \, \unit{\ampere\metre\squared}$ constant. The influence of the magnetic torque on the trajectory was again measured by the value of the $x(t)$-coordinate after $\Delta t= 1 \,\unit{\second}$ of simulation time at an inclination opening angle of $\varphi = 10^\circ$. Results are shown in Fig.~\ref{fig:f1} and Fig.~\ref{fig:f2}.
Fig.~\ref{fig:f1} shows the final $x$-deviation of the solid sphere from the straight trajectory as a function of the initial orientation of the magnetic moment and the magnetic field. It can be seen in (a) that,  when varying $\bm{\mu}_0$, two critical points, i.e., points at which the first derivative vanishes, appear in all deviation curves. For very low values of $\mu$, the absolute deviation decreases smoothly after the critical points are reached. For higher values of $\mu$, the behavior after the critical points becomes erratic. Conversely, varying $B$ does not cause erratic behavior to appear, as seen in Fig.~\ref{fig:f1}~(b). The absolute deviation generally increases with increasing $B$, but multiple pairs of critical points emerge. This behavior can be understood by looking at the selected trajectories in Fig.~\ref{fig:f2}.
Comparing Fig.~\ref{fig:f1}~(a) with Fig. \ref{fig:f2}~(a), we see that the curvature for low values of $\mu$ increases with $|\theta|$. However for $|\theta|>\frac{\pi}{2}$, the downward rolling is at first opposed by the magnetic torque reducing the total distance rolled and hence also the total deviation. Fig.~\ref{fig:f2}~(b) reveals that this is caused by the magnetic torque overcompensating the gravitational torque, causing the sphere to loop around the origin. Fig.~\ref{fig:f2} (c,d) shows that the solid sphere twists around the $x=0$ line. Depending on the azimuthal angle $\overline{\theta}$, the solid sphere crosses the line. This divides the graphs in Fig.~\ref{fig:f1}~(b) into two regions: For values of $\overline{\theta}$ between around $\frac{-\pi}{4} \, \unit{rad}$ and $\frac{\pi}{4} \, \unit{rad}$, the sphere is increasingly deflected from the straight path with increasing $|\overline{\theta}|$. At the first pair of critical points $\overline{\theta} \approx \pm \frac{\pi}{4} $, the trajectories start to curve towards the $x=0$-line again. The pair of zeros at $\overline{\theta}=\pm \frac{\pi}{2} \, \unit{rad}$ represents the point at which the path ends exactly on that line. For $|\overline{\theta}| > \frac{\pi}{2}$, the sphere crosses over $x=0$, causing the deviation curves to invert. In general, the twisting motion seen in Fig.~\ref{fig:f2}~(c,d) also becomes tighter around the $x=0$ line with increasing values of $|\overline{\theta}|$. This explains the decrease of total deviation observed for high values of $B$ in Fig.~\ref{fig:f1}~(b).\\
In Fig.~\ref{fig:g} we set $B=5 \cdot 10^{-5}\,\unit{\tesla}$, $\mu_0 = 1.0\,\unit{\ampere\metre\squared}$, $\theta=\overline{\theta}=0^\circ$, and $\overline{\psi}=0^\circ$, varying $\psi \in [-\pi,\pi)$. In addition, we repeated the simulations with the roles of $\bm{\mu}$ and $\bm{B}$ exchanged, also seen in Fig.~\ref{fig:g}. All other simulation parameters were kept from the previous simulation. Since this causes no deviation in $x(t)$, we used the value of the $y(t)$-coordinate after $\Delta t = 1\,\unit{\second}$ as measure for the influence of the magnetic torque.
Fig.~\ref{fig:g} clearly shows that for polar angles between about $\frac{-3\pi}{4}\,\unit{rad}$ and $0\,\unit{rad}$, the magnetic force fully dominates over the gravitational force, thus the solid sphere oscillates in the $yz$-plane without rolling down as seen in (b). This behaviour transitions to oscillations which are eventually overcome by gravity at a specific point in time. The more favourable the downhill motion is for the total magnetic energy, the sooner this critical point is reached. In contrast to Fig.~\ref{fig:f1}, the mirror symmetry under exchange of $\bm{\mu}$ and $\bm{B}$ is readily explained by the fact that we can write $\bm{\mu}(t)=\mu(0, \sin \psi(t), \cos \psi(t))^\top$ since $\bm{\mu}(t)$ remains always in the $yz$-plane. The magnetic torque thus reduces to
\begin{equation}
    \bm{\tau}_\text{em}(t)=\mu B\sin \psi(t) \bm{e}_x.
\end{equation}
Exchanging the roles of $\bm{B}$ and $\bm{\mu}$ simply means to reverse the angle $\psi(t)$ to $2 \pi - \psi(t)$, causing the observed sign flip of $\bm{\tau}_\text{em}$ as sinus is an odd function. The immediate difference to the conditions in Fig.~\ref{fig:f1} is that no torque breaking the mirror symmetry of the system with respect to the $yz$-plane can occur. This constrains $\bm{\mu}$ to the $yz$-plane, reducing the effective rotational degrees of freedom to one, parametrized by an antisymmetric function.

\section{CONCLUSION}
\label{sec:conclusion}
We investigated the motion of a ferromagnetic solid sphere rolling without slipping on an incline in an external magnetic field. Our simulations show that the downhill trajectory of such a sphere is considerably influenced by a magnetic field as small as that of the Earth. In general, a straight downhill trajectory acquires a transverse component similar to that of the Lorentz force. However, the deflection is highly dependent on the initial direction and magnitude of both $\bm{\mu}_0$ and $\bm{B}$. 

Surprisingly, we did not observe any symmetry on exchanging the data of $\bm{B}$ and $\bm{\mu}_0$ when investigating the influence of the azimuthal angular deviation from alignment of $\bm{B}$ and $\bm{\mu}$ despite the fact that exchanging them would only change the sign of the magnetic torque. However, a mirror symmetry emerges in the $y(t)$-coordinate when the polar angle is varied under this exchange. Moreover, we could identify three qualitatively different trajectories: Curved downward trajectories not twisting around the $x=0$ line for low values of $\mu_0$ and $B$, trajectories twisting around the $x=0$ line for high values of $B$ and high azimuthal deviations of $\bm{B}$ from $\bm{\mu}_0$, and trajectories looping around or oscillating at the origin for high values of $\mu_0$ and high values of azimuthal or polar deviation. 

To conclude, the apparent simplicity of the system paired with an intricate dependence on initial values turns it into an excellent proof-of-concept for the Lagrangian-d'Alembert approach to non-holonomic, dissipative systems. We demonstrated that even with multiple dissipative forces, the equations of motions can be solved numerically with sufficient stability, and provided a software to that end. In the future, it may be fruitful to investigate the system for chaotic behavior. Since the approach is also valid for $\varphi = 0^\circ$, i.e., a flat surface, comparison with analytical solutions available in certain such cases may be of interest.

\begin{acknowledgments}
E.Y.V., T.M., and R.C.R. acknowledge financial support provided by the Deutsche Forschungsgemeinschaft (DFG) via Project No. 514141286. 
\end{acknowledgments}

\appendix

\section{Accuracy and Friction}
\begin{figure}[htbp]
    \centering
    \includegraphics[width=\textwidth]{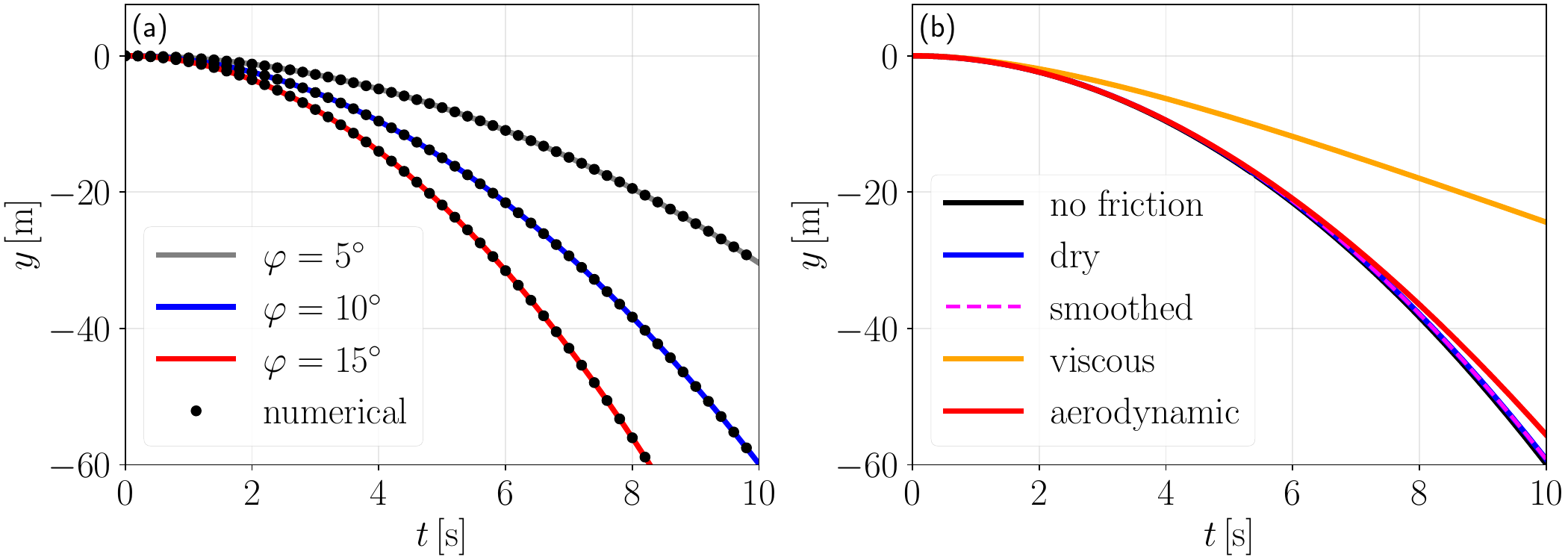}
    \caption{Numerical and analytical trajectories for different incline angles (a) and numerical trajectories for different kinds of friction (b). Fig. (a) shows no noticeable deviation between numerical and analytical values. Fig. (b) shows that the difference between discontinuous dry Coulomb friction and a smoothed variant are minimal, as is the general influence of this friction compared to no friction. Aerodynamic and viscous friction cause notable discrepancies.}
    \label{fig:a}
\end{figure}

In the absence of a magnetic field and friction, the considered system is analytically solvable. This fact was used to evaluate the accuracy of the numerical simulation and the influence of different kinds of friction. For an incline angle $\varphi$, initial positions $x_0$, $y_0$, and $z_0$ as well as initial velocities $v_{x,0}$, $v_{y,0}$, and $v_{z,0}$, the analytical motions are given by 
\begin{equation}
\begin{aligned}
    x(t) &= v_{x,0}\,t + x_0 \\
    y(t) &= -\frac{5g}{14} \sin \varphi \cos \varphi \,t^2 + v_{y,0}\,t +y_0 \\
    z(t) &= \frac{5g}{14}(1 - \cos^2 \phi) \,t^2 + v_{z,0}\,t + z_0,
\end{aligned}
\end{equation}
where $g$ is the gravitational acceleration of Earth. We compared these equations for chosen values of $\varphi$ in Fig.~\ref{fig:a}~(a). Furthermore, we compared how different kinds of friction influenced a system under identical initial values in Fig.~\ref{fig:a}~(b). The compared models were dry Coulomb friction and aerodynamic drag as in Eq.~\ref{eq:friction}, the smoothed Coulomb variant under the approximation of Eq. \ref{eq:smoothfriction}, and a viscous approximation of Coulomb friction given by 
\begin{equation}
F_\text{e} = - \nu_\text{r} \dot{\bm{r}}_\perp.
\end{equation}
As shown, the simulation agrees closely with the analytical solution. The maximal discrepancy between analytical and numerical values is $60.8 \, \unit{\micro\metre}$ for $\varphi = 5^\circ$, $120 \, \unit{\micro\metre}$ for $\varphi=10^\circ$, and $175 \, \unit{\micro\metre}$ for $\varphi=15 ^\circ$, thus not exceeding the micrometer scale. The remarkable discrepancy of the viscous approximation let us to discard this approach, while the smoothed Coulomb friction seemed sufficient to approximate the discontinuous dry friction. The almost negligible influence of Coulomb friction is in accordance with theoretical predictions in the case of ideal rolling without slipping.

\end{document}